\documentclass[%
 reprint, 
 amsmath,amssymb,
 aps, 
]{revtex4-2}

\usepackage{subcaption}
\usepackage{siunitx}
\usepackage[utf8]{inputenc}
\usepackage[T1]{fontenc}
\usepackage{pgf}
\usepackage{comment}
\usepackage{graphicx}
\usepackage{dcolumn}
\usepackage{bm}
\usepackage{amsfonts,amsmath,amsthm}
\usepackage{xcolor}
\usepackage{ulem}

\begin{document}

\preprint{APS/123-QED}

\newcommand{\vkeff}{\vec{k}_{\text{eff}}}
\newcommand{\keff}{k_{\text{eff}}}
\newcommand{\red}[1]{\textcolor{red}{#1}}
\newcommand{\blue}[1]{\textcolor{blue}{#1}}

\title{\textbf{Impact of Raman lasers beam profile inhomogeneities on a spaceborne quantum accelerometer} 
}%

\author{Benoit Kaczmarczuk}%
\author{Gabriel Ducasse}%
\author{Louis Pagot}%
\author{Quentin Beaufils}%
\author{Franck Pereira dos Santos}%

\email{Contact author: franck.pereira@obspm.fr}
\affiliation{%
  LTE, Observatoire de Paris - Université PSL, Sorbonne Université, Université de Lille, LNE, CNRS, Paris, France
}%

\date{\today}

\begin{abstract}
We report on the study of the impact of laser beam profile inhomogeneities on the acceleration measurements of a quantum accelerometer in space. This sensor uses a low ballistic expansion source based on delta-kick collimated Bose-Einstein condensates, and an interferometer configuration based on a sequence of Raman double diffraction beamsplitters, best suited to the microgravity environment. We characterize both phase and intensity fluctuations in terms of power spectral densities and calculate their impact on the interferometer phase, averaging over all trajectories of the atoms, via numerical simulations based on randomly drawn laser profiles as well as via analytical treatments. We show that high quality optics, such as those realized for gravitational wave detectors, combined with large initial size atomic sources, will allow to reduce interferometer phase bias and fluctuations to the order of $1$~mrad corresponding to accelerations in the low $10^{-12}\text{m.s}^{-2}$ range.
\end{abstract}

\maketitle

 \section{\label{sec:Introduction} Introduction\protect\\}

Quantum accelerometry has emerged as a promising technology for applications in navigation, geodesy, and fundamental physics. The space environment offers unique advantages for such instruments, including extended free-fall times and reduced environmental noise, which significantly enhance measurement precision. Beamsplitters based on light pulses allow for placing atoms in superposition of well defined external states, with well controlled momentum separation for free falling atoms \cite{Borde1989,Moler1992}. 
This is a key feature for the realization of atom interferometers with stable scale factors \cite{Kasevich1991,Riehle1991, Rasel1995}, which find numerous applications in inertial sensing and fundamental physics \cite{Geiger2020}. Conversely, imperfections in the laser beam profiles limit the accuracy and long term stability of the measurements \cite{Fils2005,Gauguet2009,Louchet2011,Bade2018}, and the impact of phase and intensity distortions have been the subject of numerous theoretical \cite{Wicht2005,Hogan2011,Cervantes2024,Pagot2025,Seckmeyer2025,Mouelle2025,Pagot2026} and experimental investigations \cite{Schkolnik2015,Zhou2016,Trimeche2017,Karcher2018,Zhang2021,Xu2024,Gaudout2025,Luo2025,Junca2026}.

In this paper, we study the impact of the optical quality, in terms of phase and intensity profiles, of the Raman laser beams and retroreflecting optics, on the contrast and systematics of a space-borne cold atom interferometer. We consider the case of a double diffraction Raman interferometer \cite{Leveque2009,Malossi2010,Zhou2015}, a configuration of reference for quantum accelerometry in microgravity \cite{Aguilera2014,Carraz2014,Trimeche2019,Leveque2023}. Rather than characterizing optical defects via a modal decomposition on a finite Zernike polynomial basis \cite{Pagot2025}, we use a spectral characterization \cite{Hogan2011} of the phase, intensity or mirror height profiles, allowing to cover all the relevant spatial frequency domain, going further than the low frequency usually well captured by a finite set of the first Zernike polynomials, such as provided by wavefront sensors. The high frequency domain, usually characterized by an overall figure such as rms fluctuations, is better described by spectra based on high spatial resolution measurements, to the cost of restraining the study to a limited surface. 
For that, we define the incoming Raman beam profile by a random draw of the phase and intensity and propagate it via Fast Fourier Transform (FFT) methods along the atom interferometer to obtain profiles at the positions of interaction with the atoms. The retro-reflecting mirror is also defined by a random draw of height value. For simplicity, we neglect the finite transverse size of the Gaussian laser beam and consider a flat intensity profile across the interaction region. This is a legitimate approximation in the context of instrument concepts based on the use of Bose-Einstein condensates in microgravity \cite{Muntinga2013}, where the size of the cloud (of a few hundred micrometers after a few seconds \cite{Leveque2023}, or a few mm after tens of seconds \cite{Struckmann2024}) remains, even after expansion, significantly smaller that the beam waist (expected to range from few mms to few cms).

\section{\label{sec:} Methodology\protect\\}

\subsection{Interferometer configuration}

The considered interferometer consists in a sequence of three Raman pulses, separated by free evolution times $T$. In this configuration (as well as in more standard accelerometer configurations), the two Raman lasers are overlapped by injection in a common mono-mode optical fiber, before being sent to the atoms and retroreflected on the reference mirror. The lasers have wavevectors $k_1$ and $k_2$ on their way forth, and $k'_1$ and $k'_2$ on their way back. We denote $L$ the distance between the atoms at the first pulse of the interferometer and the mirror, and $2l$ the separation between the two wavepackets at the middle mirror pulse. 

\begin{figure}[h]
  \centering
  \includegraphics[width=8.5cm]{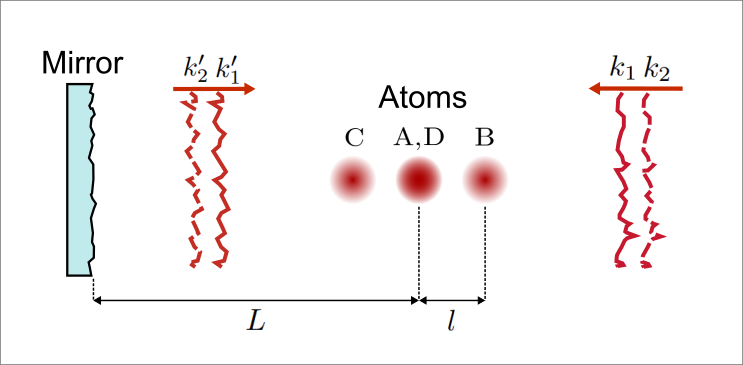}\\
  \caption{Double diffraction interferometer scheme and Raman laser configuration. The atomic positions are displayed for the first ($A$), second ($B$ and $C$) and third ($D$) pulse. The two incident Raman laser beams (on the right) are supposed to be the same. Their wavefronts evolve due to free space propagation and to the retro-reflection on the imperfect mirror.}\label{interferophase}
\end{figure}

\subsection{Atom-light interaction}

The Raman field propagation modifies both the wavefront and the intensity profile of the beam, which in turn affect the atom interferometer via direct impression of the phase for the former, and AC Stark shift for the latter. In the following, we consider both effects. 

\subsubsection{Phase shift}

The phase shift along the right atomic path is given by:
\begin{equation}
\begin{aligned}
\Phi_I&=-(\phi'_1(A)-\phi_2(A))+(\phi'_1(B)-\phi_2(B))\\
&-(\phi_1(B)-\phi'_2(B))+(\phi_1(D)-\phi'_2(D))
\end{aligned}
\end{equation}

where $\phi_i$ (resp. $\phi'_i$) is the phase of the laser $k_i$ (resp. $k'_i$), and A, B and D are the positions of the right wavepacket at the three pulses.
The phase shift along the left path is:
\begin{equation}
\begin{aligned}
\Phi_{II}&=-(\phi_1(A)-\phi'_2(A))+(\phi_1(C)-\phi'_2(C))\\
&-(\phi'_1(C)-\phi_2(C))+(\phi'_1(D)-\phi_2(D))
\end{aligned}
\end{equation}
where C is the position of the left wavepacket at the mirror pulse.

Using these relations, the interferometer phase is
\begin{equation}
\Phi=\Phi_{II}-\Phi_{I} \label{eq:Phi}\\
\end{equation}

\subsubsection{AC Stark shifts}

Intensity inhomogeneities in the beam lead to a  phase contribution related to the difference in the light shifts (AC Stark shifts) experienced by the spatially separated wavepackets at the mirror pulse \cite{Pesche2026}.

The atomic wavepackets experience the intensities of four laser beams, two incoming and two retroreflected, which add up when interference effects are avoided by a proper choice of the Raman laser polarizations. The resulting light shifts $\delta E$ and $\delta F$ on the two different hyperfine states $F=2$ and $F=1$ are thus given by a linear combination of these four laser intensities, and are different for the two paths. 

These light shifts are given by
\begin{equation}
\begin{aligned}
\delta{E_{I}} &= \alpha(I_1(z_{Ci}) + I_1(z_{Cr}))+ \beta( I_2(z_{Ci}) + I_2(z_{Cr}))\\
\delta{F_{I}} &= \gamma(I_1(z_{Ci}) + I_1(z_{Cr}))+ \delta( I_2(z_{Ci}) + I_2(z_{Cr}))\\
\delta{E_{II}} & = \alpha(I_1(z_{Bi}) + I_1(z_{Br}))+ \beta( I_2(z_{Bi}) + I_2(z_{Br}))\\
\delta{F_{II}} &= \gamma(I_1(z_{Bi}) + I_1(z_{Br}))+ \delta( I_2(z_{Bi}) + I_2(z_{Br}))
\end{aligned}
\end{equation}
where $I_1$ and $I_2$ are the normalized intensities and the parameters $\alpha,\beta,\gamma,\delta$ depend on the intensities of the Raman lasers, and their detuning with respect to the D2 line. For a detuning of $\Delta/2\pi=-1$ GHz, a Rabi frequency of 16.5 kHz and a ratio between the laser intensities that compensates the differential ligh shift, we have $(\alpha=-13.1, \beta=-3.1, \gamma=2.5, \delta=-18.7) \times 2\pi $ krad/s.

The different light shifts from the mirror pulse at the two wavepacket positions leads to an interferometer phase shift given by :
\begin{equation}
\Phi_{LS}=\delta LS \cdot \tau_M
\label{eq:phils}
\end{equation}
with $\tau_M$ the duration of the mirror pulse, and $\delta LS$ an effective light shift difference given for the double diffraction interferometer by
\begin{equation}
\delta LS = 3(\delta{F_{I}} - \delta{F_{II}})/4 + (\delta{E_{I}} - \delta{E_{II}})/4
\label{eq:ls}
\end{equation}

\subsection{Light field propagation}

In order to evaluate the phase and intensity profile of both Raman light fields at the atom-light interaction positions, we start by generating a laser beam profile in an initial plane, with either phase or intensity fluctuations, over a grid of $5000 \times 5000$, with a pixel size of $10 \times 10\mu$m. With this sampling and the $780$~nm wavelength of the laser beams, the fields can be propagated by solving the Helmholtz equation through FFT calculation over a typical distance of $60$~cm without aliasing \cite{Kozacki2008}. This $60$~cm distance is longer than the maximal propagation length of $36$~cm used to obtain the results presented in the following.  

We propagate this field, and calculate the transverse laser field profiles at the different positions where atom-light interactions take place. We then calculate the mean value $\overline{\Phi}$ and standard deviation $\sigma_\Phi$ of the interferometer phase shift from the linear combination of the local laser phases at the pulses position, according to equation~\ref{eq:Phi}, by randomly sampling the initial positions and velocities of the atoms in the transverse plane. The initial positions of the atoms are drawn from a Gaussian distribution, with a standard deviation chosen from the interval [$10$-$300$]~µm. The velocity distribution is also Gaussian with a standard deviation corresponding to a fixed temperature of $100$~pK, which can be obtained using delta kick collimation methods~\cite{Ammann1997,kovachyMatterWaveLensing2015,Gaaloul2022,Deppner2021}. 

We also evaluate the impact of spatial inhomogeneities of light intensity using equation~\ref{eq:phils} by calculating the mean value $\overline{\Phi}_{LS}$ and standard deviation $\sigma_{\Phi_{LS}}$ of the AC Stark shift.

These calculations are repeated for several randomly generated profiles corresponding to a given model spectrum. Averaging results over a large number of profiles, we expect the mean contribution to the interferometer phase to tend to zero. Thus, we evaluate the impact of these two effects through their standard deviations $\Sigma_{\overline{\Phi}}$ and $\Sigma_{\overline{\Phi}_{LS}}$. Those quantities can be considered as characteristic biases associated to randomly drawn inhomogeneous phase and intensity profiles.

In order to evaluate the impact of inhomogeneities on the contrast \cite{Pesche2026}, we also calculate the mean of the phase dispersions within the cloud, which are similar from one profile to another, $\langle \sigma_\Phi \rangle$ and $\langle \sigma_{\Phi_{LS}} \rangle$. 

\subsection{Validation with model spectra}

%
In this section, in order to validate the methodology, we evaluate with both Monte-Carlo simulations and analytical expressions, the impact of the surface of optics on the standard deviation of the interferometer phase shift $\Sigma_{\overline{\Phi}}$. For simplicity it is attributed to a single optical element, the retroreflecting mirror which is defined by a model spectra.

\begin{figure}
    \centering
    \includegraphics[width=7cm]{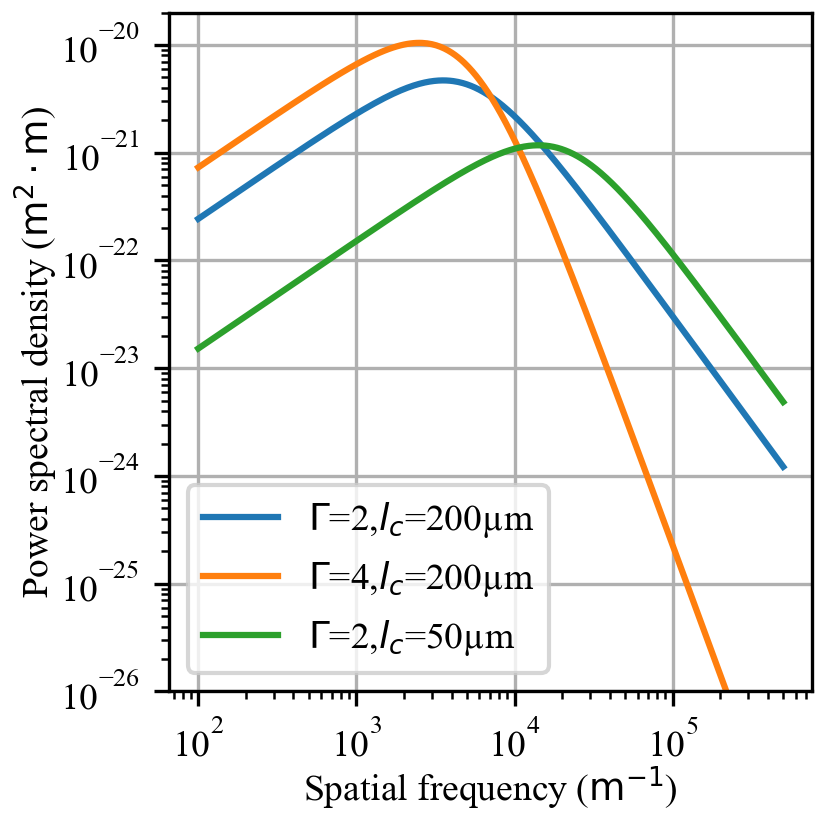}
    \caption{Pseudo1D~PSDs (eq.~\ref{eq:PSD_pseudo1D}) for model PSDs of the surface texture for different characteristic lengths $l_c=2\pi/k_c$ and order parameters $\Gamma$ (eq.~\ref{eq:PSD_models}).}
    \label{fig:spectrumABC}
\end{figure}

To grasp the relative influence of phase inhomogeneities depending on their characteristic frequency, we start by investigating the case of a model PSD of the surface texture, corresponding to a low-pass filtered white noise spectrum $S(k_x,k_y)=S_0$. The filter~\cite{Church1988} is characterized by a cutoff angular spatial frequency $k_c$ (and corresponding characteristic length $l_c=2\pi/k_c$) and an order parameter $\Gamma$: 
\begin{equation}\label{eq:PSD_models}
H(k)=\frac{1}{(1+k^2/k_c^2)^{(\Gamma+1)/2}}
\end{equation}
with $k=2\pi f$ the angular spatial frequency. The pseudo1D PSD~\cite{Jacobs_PSD_2017} of mirror surface texture is defined as
\begin{equation} \label{eq:PSD_pseudo1D}
S_h(k)=k H(k) S_0
\end{equation}
Figure \ref{fig:spectrumABC} illustrates the pseudo1D~PSD for different characteristic lengths and filter parameters. The rms $\sigma_h$ of the mirror surface texture is
\begin{equation}
\sigma_h^2=\frac{1}{2\pi}\int S_h(k)dk
\end{equation}

\begin{figure}[h]
  \centering
  \includegraphics[width=7cm]{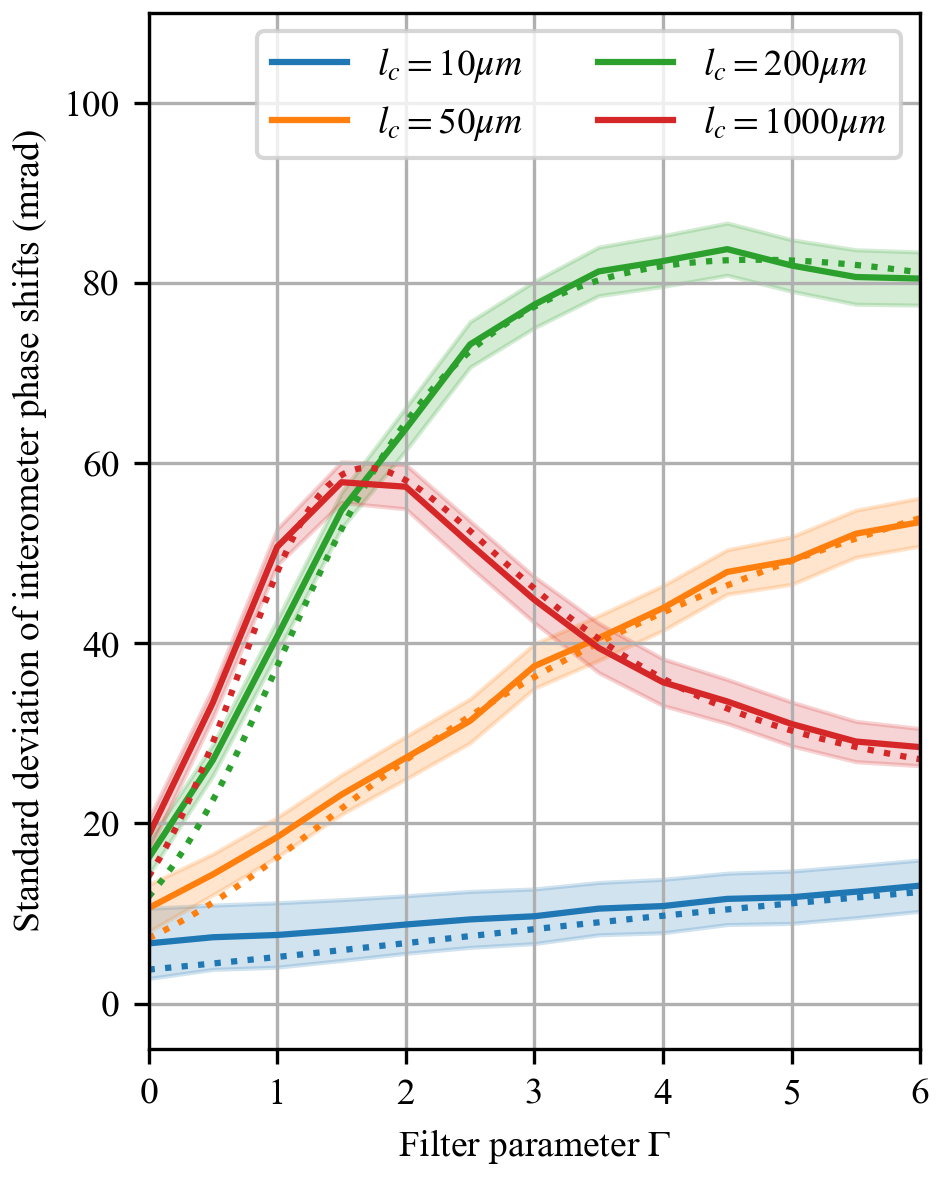}\\
  \caption{
  Standard deviation of the interferometer phase shift $\Sigma_{\overline{\Phi}}$ as a function of the order parameter $\Gamma$ for different characteristic lengths $l_{c}$. Solid line : results of the Monte-Carlo simulations, dashed lines : analytical expressions (see section \ref{subsecappmirror}). }\label{fig:phivsC}
\end{figure}

Figure \ref{fig:phivsC} displays $\Sigma_{\overline{\Phi}}$ as a function of the order parameter $\Gamma$ for different characteristic lengths $l_c$, obtained from simulations of an interferometer with total duration $2T=2$~s and an initial cloud size $\sigma_0=30~\mu$m. For each pair of parameters $(\Gamma,l_c)$, the amplitude $S_0$ is adjusted to keep the rms of profile height deviations at a fixed value of $\sigma_h=\lambda/100$.

For low values of $l_c$, we observe a monotonous increase of the phase shift standard deviation $\Sigma_{\overline{\Phi}}$ with the order of the filter, up to a maximum value of about 80 mrad. $\Sigma_{\overline{\Phi}}$ also increases with the characteristic length $l_{c}$, except for the value $l_c=1000 \mu$m, for which we observe a non-monotonous behavior.

The different curves in figure \ref{fig:phivsC} can be interpreted by comparing the characteristic sizes of the beam and the atomic cloud. For instance for the $100$~pK atomic cloud temperature considered here, the transverse size of the cloud is $\sigma_2=100~\mu$m (resp. $\sigma_3=200~\mu$m) at the second (third) pulse. Thus, the contributions of aberrations with typical transverse size $2\pi/k$ smaller than $\sigma_2$ are strongly averaged during the last two pulses. For $l_c \ll \sigma_2$, slight increase of the order parameter $\Gamma$ amounts to redistributing contributions toward lower spatial frequencies, as shown in figure~\ref{fig:spectrumABC}, that are still averaged out over the last two pulses but have higher contribution during the first beamsplitter.


The results are in good agreement with the analytical expression given by equation \ref{eq:sigmaphi}, provided it is adapted to the case of the retroreflecting mirror, as explained in section \ref{subsecappmirror}, and with integration limits set by the size of the picture on one hand, and by the pixel size on the other hand. The analytical results are displayed as dotted lines on figure \ref{fig:phivsC}.

\section{Impact of high quality retroreflecting mirrors}

\begin{figure}
    \centering
    \includegraphics[width=7cm]{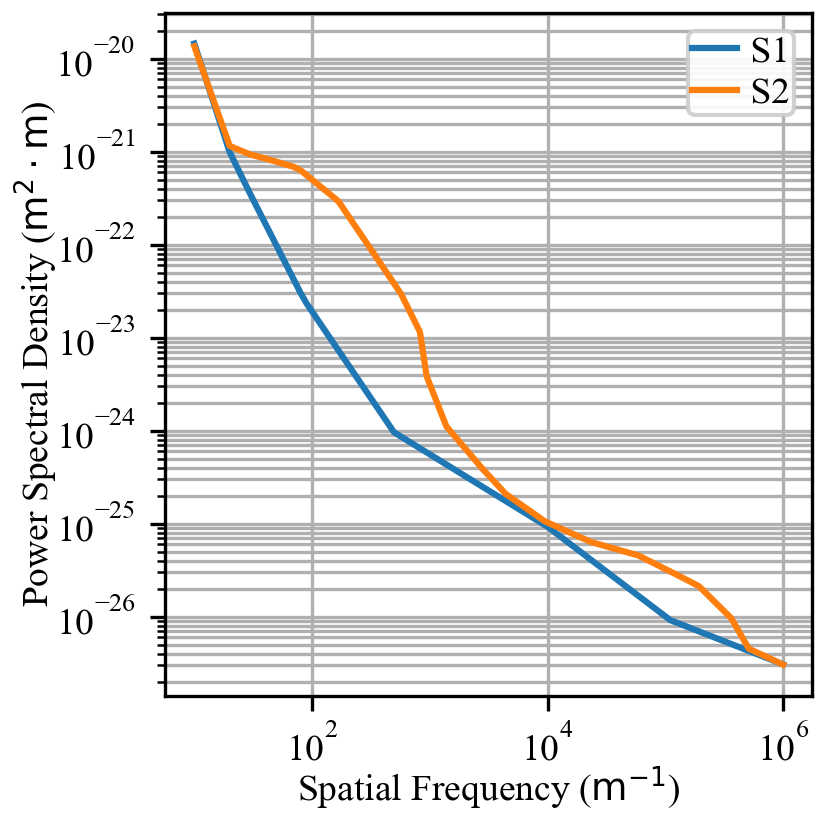}
    \caption{Models for the PSDs of surface texture of S1 (blue) and S2 surfaces (orange), derived from \cite{Hirose2014}}
    \label{fig:RealisticSpectrum}
\end{figure}

In this section, we evaluate the impact of imperfect flatness of realistic ultra high flatness dielectric mirrors, such as the ones used as test masses for gravitational wave detection. In particular, we extract two spectra out of \cite{Hirose2014}, one corresponding to the fused silica substrate of LIGO's cavity mirrors (surface S1) and the second one to coated sapphire mirrors for KAGRA (surface S2). The corresponding (approximated) spectra are displayed respectively as blue and orange lines on figure \ref{fig:RealisticSpectrum}.  The root mean square (RMS) surface roughness of S1 is $0.26$~nm, while that of S2 is $0.43$~nm integrated over the range $0.01$~mm$^{-1}$ to $1\,000$~mm$^{-1}$.


\begin{figure}[h]
  \centering
  \includegraphics[width=7cm]{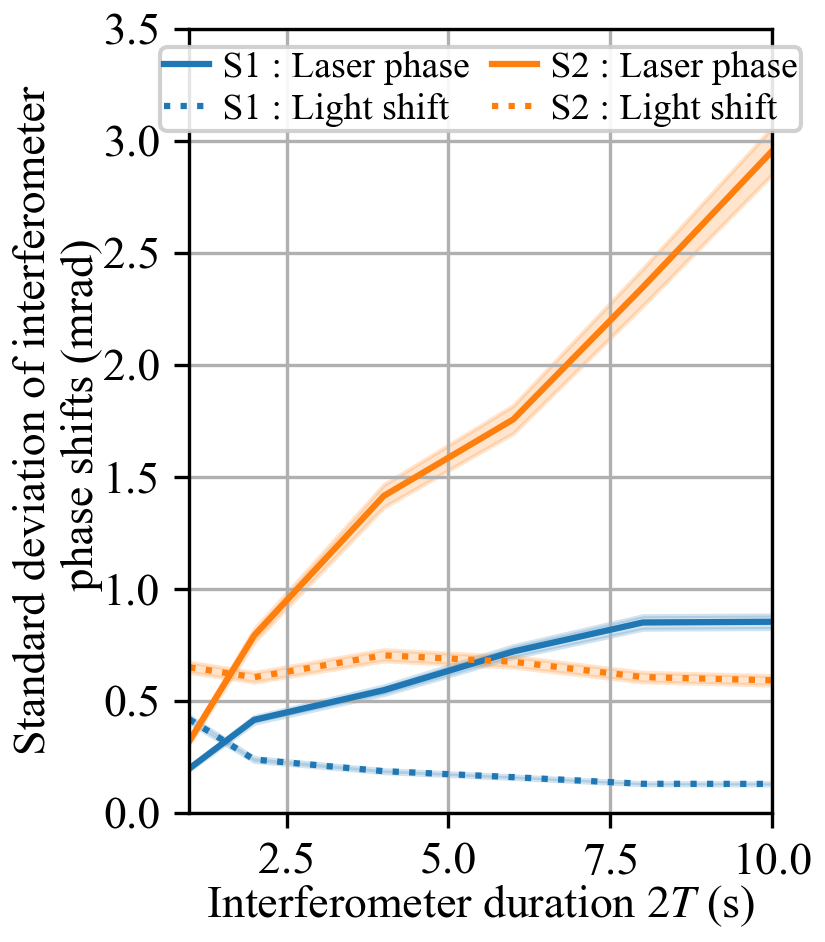}\\
  \caption{Standard deviations of interferometer phase shifts, due to laser phase fluctuations $\Sigma_{\overline{\Phi}}$ (thick lines) and light shift fluctuations $\Sigma_{\overline{\Phi}_{LS}}$(dotted lines), as a function of the interferometer time $2T$. The blue trace corresponds to the surface S1 and the orange ones to S2.}\label{fig:phirealistic}
\end{figure}

Figure \ref{fig:phirealistic} displays the characteristic phase shifts $\Sigma_{\overline{\Phi}}$ and $\Sigma_{\overline{\Phi}_{LS}}$ as a function of the interferometer duration $2T$, for a cloud size $\sigma_0=30~\mu$m. The impact of laser phase inhomogeneities increases with $2T$ for the two mirror surfaces, due to the increase of the distance between the wavepackets at the middle pulse. On the contrary, the effect of intensity inhomogeneities is found to decrease with $2T$. Note that despite this difference in trend, the effects on the interferometer phase of laser phase and intensity fluctuations are of the same order of magnitude, both in the mrad range. 

\begin{figure}[h]
  \centering
  \includegraphics[width=7cm]{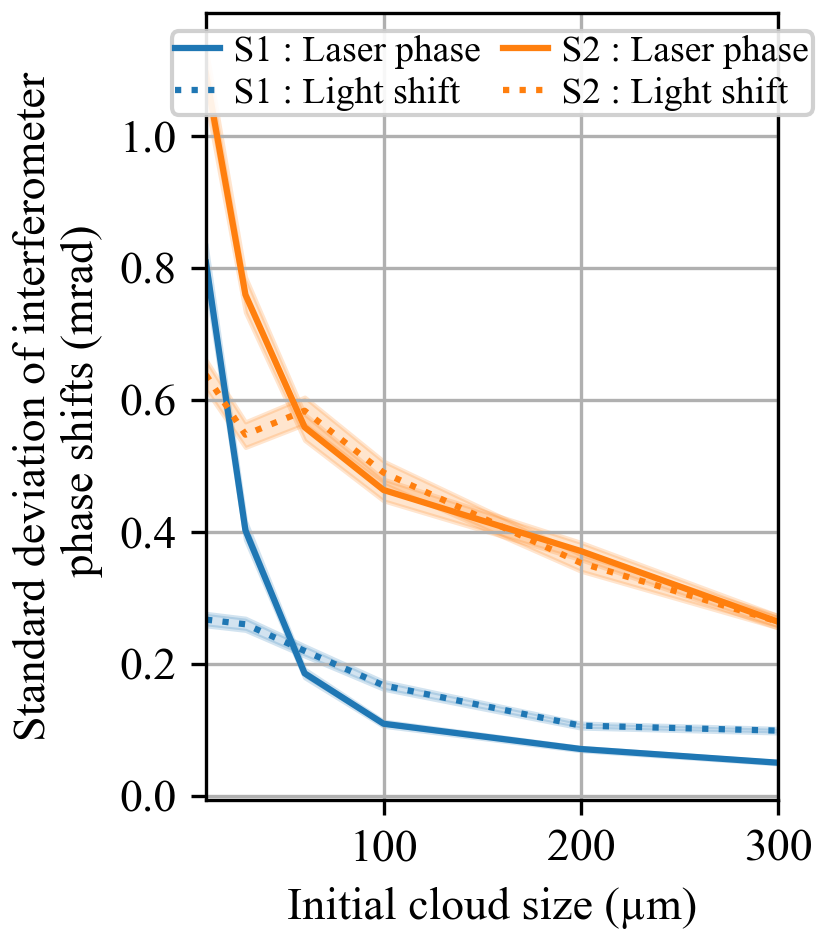}\\
  \caption{Standard deviations of interferometer phase shifts, due to laser phase $\Sigma_{\overline{\Phi}}$ (thick lines) and light shift spatial noise $\Sigma_{\overline{\Phi}_{LS}}$(dotted lines), as a function of the initial size of the atomic cloud $\sigma_0$, for an interferometer duration of $2T=2$s. The blue trace corresponds to the surface S1 and the orange ones to S2.}\label{fig:phivssize}
\end{figure}

We repeat these calculations for different cloud sizes $\sigma_0$ and for $2T=2$s. The results, displayed on figure \ref{fig:phivssize}, show the decrease of the effects of laser phase and light shift inhomogeneities when increasing the size of the atom cloud, due to spatial averaging.

\begin{figure}[h]
  \centering
  \includegraphics[width=7cm]{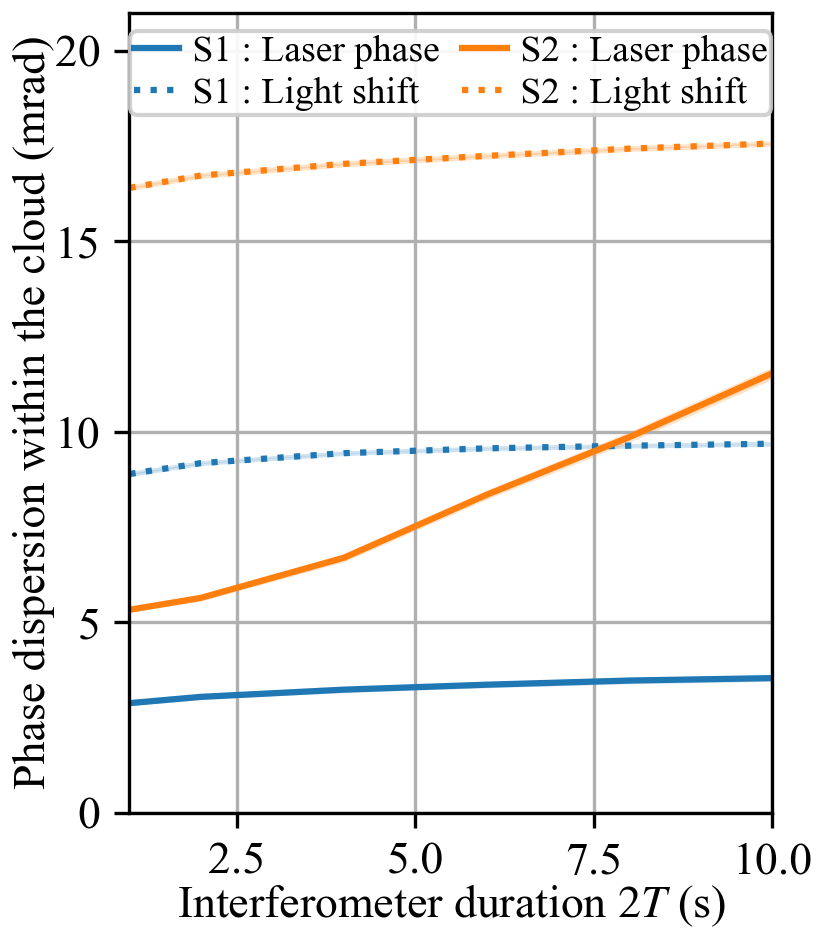}\\
  \caption{Dispersions of the phase contributions to the interferometer phase within the atomic source, $\langle \sigma_\Phi \rangle$ and $\langle \sigma_{\Phi_{LS}} \rangle$, as a function of the interferometer time $2T$. The initial size of the atomic cloud is $\sigma_0=30\mu$m. The blue traces correspond to the surface S1 and the orange ones to S2. 
 }\label{fig:disp}
\end{figure}

Finally, figure \ref{fig:disp} displays the corresponding values of $\langle \sigma_\Phi \rangle$ and $\langle \sigma_{\Phi_{LS}} \rangle$ as a function of the interferometer duration $2T$, for a cloud size $\sigma_0=30\mu$m. Even at their maximum values, their impact on the contrast, given by $\mathcal{C}=\mathcal{C}_0e^{-\langle\sigma_\Phi\rangle^2/2}$ and $\mathcal{C}=\mathcal{C}_0e^{-\langle\sigma_{\Phi_{LS}}\rangle^2/2}$, are found to be negligible. 

The analytical results agree well with the numerical simulations, provided integrals are calculated over the frequency band addressed by the simulation, with cutoffs set by the size of the picture at low frequency and by the pixel size at high frequency. 

\section{\label{sec:xx}Impact of the optical quality of the incoming beam}

In this section, we evaluate the impact of spatial phase noise in the incoming beam, which is delivered to the atoms via an optical system that would for instance collimate the output of an optical fiber.

We will use as a model for the wavefront of the incoming beam a PSD related to that of S1, considering that phase inhomogeneity in the incoming beams $\phi_i$ originates from height or thickness fluctuations in the delivery optics, neglecting index inhomogeneities, so that $\phi_i \sim k_i h$. We apply the incoming phase profile on a plane located at a distance of 10~cm from the atoms (on their right in figure \ref{interferophase}).

\begin{figure}[h]
  \centering
  \includegraphics[width=8cm]{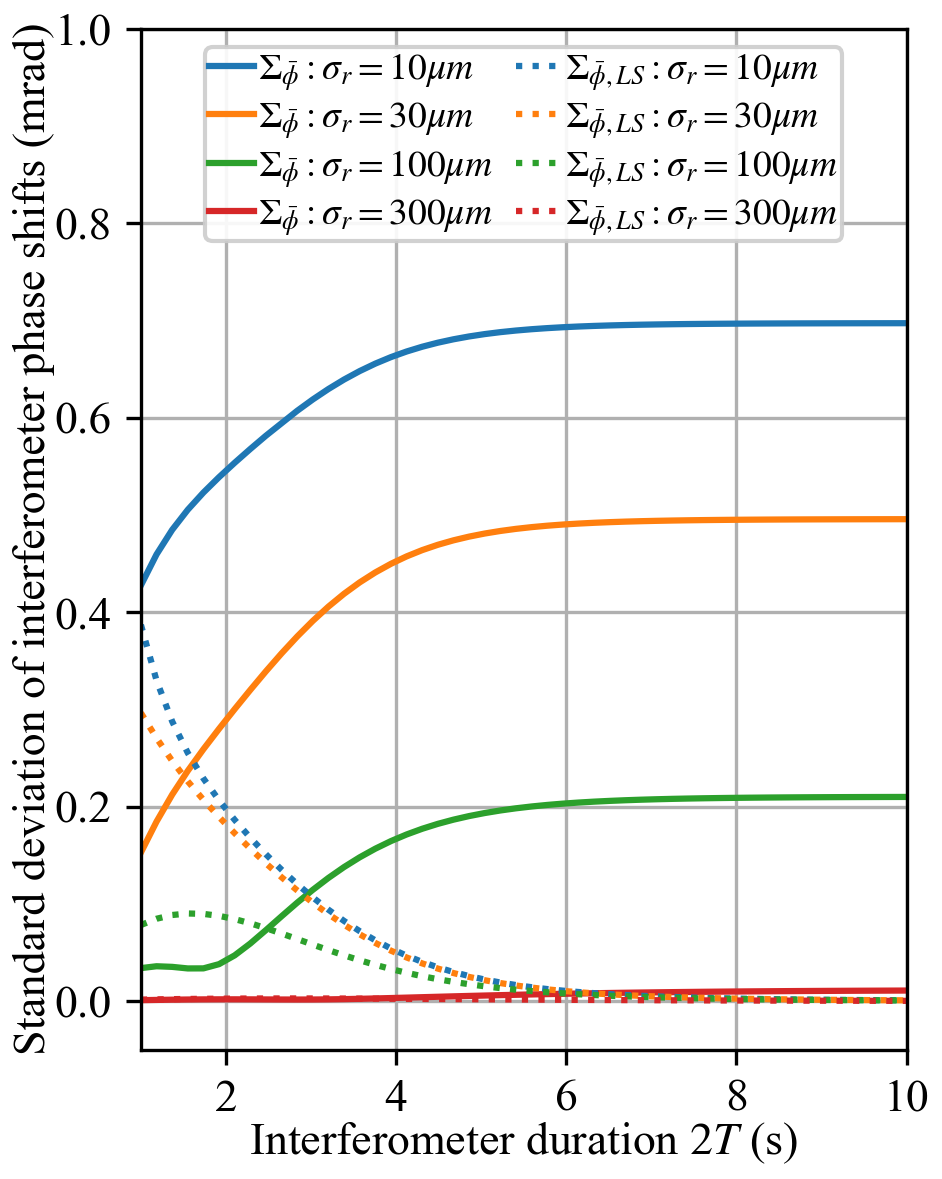}\\
  \caption{Standard deviations of interferometer phase shifts, due to laser phase  $\Sigma_{\overline{\Phi}}$ (thick lines) and light shift spatial noise $\Sigma_{\overline{\Phi}_{LS}}$(dotted lines), as a function of the interferometer duration $2T$, for different initial sizes of the atomic cloud $\sigma_r$ ($10$, $30$, $100$, and $300\ \mu$m). Phase fluctuations in the incoming beam are derived from height fluctuations of the surface S1.}
  \label{fig:phase_vs_2T_sigma}
\end{figure}

Figure \ref{fig:phase_vs_2T_sigma} displays the corresponding $\Sigma_{\overline{\Phi}}$ (thick lines) and $\Sigma_{\overline{\Phi}_{LS}}$ (dotted lines) as a function of the interferometer duration $2T$ and for the different sizes of $\sigma_0=10, 30, 100$ and $300~\mu$m. For these evaluations, we have used the analytical expressions \ref{eq:sigmaphi} and \ref{eq:sigmaphils} with the integration covering the whole frequency range [0.01-1000] mm$^{-1}$. We find effects similar in amplitude and behavior as the ones for the retroreflecting mirror.

\begin{figure}[h]
  \centering
  \includegraphics[width=8cm]{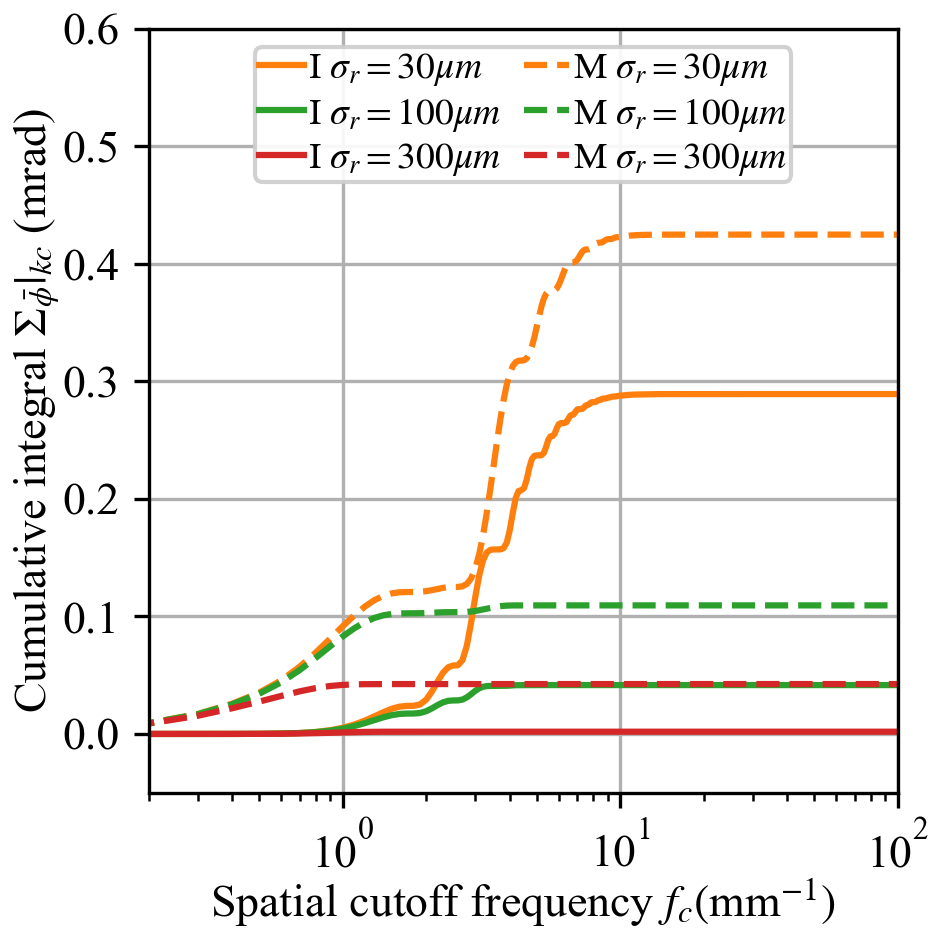}\\
  \caption{Cumulative integral of $\Sigma_{\bar{\phi}}|_{kc}$ as a function of the spatial cutoff frequency $f_c=k_c/2\pi$, for three cloud sizes $\sigma_r = 30\ \mu$m (orange), $100\ \mu$m (green), and $300\ \mu$m (red). Solid lines (I) correspond to the incoming beam, and dashed lines (M) to the mirror surface.}\label{fig:cumulative_phase}
\end{figure}

To highlight the difference between the two situations (mirror vs incoming beam), we plot in figure \ref{fig:cumulative_phase} the cumulative integrals of $\Sigma_{\bar{\phi}}$ as a function of the spatial cutoff frequency $f_c=k_c/2\pi$, for $2T=2$s and for different values of $\sigma$. In both cases, we find that the dominant contributions to the results lie in the [0.1-10] mm$^{-1}$ band, with the averaging over the size acting as a low pass filter, truncating the contribution of high frequencies. A notable difference between the two cases lies in the relative contribution of fluctuations at low spatial frequencies ($f < 1~\text{mm}^{-1}$), with a larger impact for the mirror than for the incoming beam. This relates to the efficient common-mode suppression of the lowest frequency components in the difference between the back and forth beams.

Finally, the impact of intensity inhomogeneities in the incoming beam can be evaluated in a similar manner, as mentioned in the appendix, provided the knowledge of their PSD.


\section{Impact for space accelerometry}
\label{sec:}

Our analysis shows that phase biases related to intensity and phase spatial noise in the Raman lasers, as well as their variations when accounting for typical time fluctuations in the initial position and velocity of the atomic source, can be reduced down to the mrad level, if not better, by combining feasible ultra high quality optical elements and sufficiently spatially extended atomic sources. For an interferometer duration of $2T=10~$s, this translates into an acceleration stability and accuracy in the low $10^{-12}\text{m.s}^{-2}$ range, of relevance for applications in space geodesy, as proposed and studied for instance in \cite{Carraz2014,douch2018,Trimeche2019,leveque2019,HosseiniArani2024,mu2024,zingerle2024,stray2025}.

\section{Conclusion}

We evaluated the impact of phase and intensity fluctuations in the laser beams onto the interferometer phase and contrast, accounting for the propagation of the laser fields and for the dispersion of the atomic trajectories via Monte-Carlo simulations. The results are found to agree well with analytical expressions. 
Our study highlights the role played by the phase and intensity noise at high spatial frequencies, whose impact can be mitigated by increasing the initial size of the cloud. Finally, we show that the optical flatness of high quality dielectric mirrors, such as developed for gravitational wave detectors, would allow to comply with a requirement on the bias on a space accelerometer in the low $10^{-12}$ m/s$^2$ range.

Our approach allows specifying requirements for the optical quality of the laser beams, over a broader spatial frequency range than the usual basis of low order Zernike polynomials used to decompose laser or optics aberrations. For instance, templates for the PSDs of phase and intensity fluctuations of the laser beams can be defined, for which the probability for the phase bias to be lower than a certain target value would be larger than a given threshold (e.g. 95\%). Though we have performed this study in the context of accelerometry in space, with the specific geometry of a double-diffraction interferometer, it can readily be adapted to atom interferometers based on other architectures, such as based on single diffraction and/or alternative sequences of Raman pulses.


\begin{acknowledgments}

This work benefited from a government grant managed by the Agence Nationale de la Recherche under the Plan France 2030 with the reference “ANR-22-PETQ-0005” (QAFCA project), and was partially funded by the European Union's Horizon Europe research and innovation programme under grant agreements No 101189541 (CARIOQA‐PHB project). We also acknowledge support from CNES (project DS/DAP/EOT-2024.0009975).

\end{acknowledgments}

\section{Appendix}

We provide in this appendix analytical expressions for the standard deviations of the mean phase $\Sigma_{\overline{\Phi}}$ and the mean phase shift induced by light shifts $\Sigma_{\overline{\Phi}_{LS}}$. We consider different cases, of an incoming laser beam with phase fluctuations only, or with intensity fluctuations only, or of a mirror with surface height fluctuations.  

\subsection{Impact of phase fluctuations}

For an atom with initial transverse positions ${x_0,y_0}$ and velocities ${v_x,v_y}$, the laser phase at the atom's position at time $t$ can be written as 
\begin{equation}
\phi(x_1,y_1,z_1)=\iint\delta(x-x_1)\delta(y-y_1)\phi(x,y,z_1)dxdy
\end{equation}
with $x_1=x_0+v_xt$ and $y_1=y_0+v_yt$.

Introducing $\hat{\phi}$ the Fourier transform of $\phi$, the laser phase writes as
\begin{equation}
\begin{aligned}
\phi(x_1,y_1,z_1)&=\frac{1}{4\pi^2}\iiiint\delta(x-x_1)\delta(y-y_1)e^{ik_xx}e^{ik_yy}\\
&\times\hat{\phi}(k_x,k_y,z_1)dxdydk_xdk_y
\end{aligned}
\label{eq:int4}
\end{equation}
From the paraxial equation describing the free space propagation of the laser field, we get
\begin{equation}
\hat{\phi}(k_x,k_y,z_1)=\hat{\phi}(k_x,k_y,0)\cos \left(\frac{(k_x^2+k_y^2) \lambda  z_1}{ 4\pi }\right)
\label{eq:fft}
\end{equation}
Injecting \ref{eq:fft} into \ref{eq:int4}, and going back to real space for $\phi$, we get
\begin{equation}
\phi(x_1,y_1,z_1)=\iint h(x,y)\phi(x,y,0)dxdy
\end{equation}
with
\begin{equation}
\begin{aligned}
h(x,y)=&\frac{1}{4\pi^2}\iint e^{-ik_xx_1}e^{-ik_yy_1}\cos \left(\frac{(k_x^2+k_y^2)\lambda  z_1}{4 \pi }\right)\\
&\times e^{ik_xx}e^{ik_yy} dk_xdk_y
\end{aligned}
\end{equation}
The Fourier transform $\hat{h}$ of $h$ is thus given by 
\begin{equation}
\hat{h}(k_x,k_y)=e^{-ik_xx_1}e^{-ik_yy_1}\cos \left(\frac{(k_x^2+k_y^2)\lambda  z_1}{4 \pi }\right)
\end{equation}

Averaging the laser phase over gaussian velocity and position distributions characterized by a rms size $\sigma_0$ and rms velocity $\sigma_v$, we get
\begin{equation}
\bar{\phi}(z)=\iint \bar{h}(x,y,z)\phi(x,y,0)dxdy
\end{equation}
with
\begin{equation}
\hat{\bar{h}}(k_x,k_y)=e^{-\frac{1}{2}(k_x^2+k_y^2)\sigma(t)^2}\cos \left(\frac{(k_x^2+k_y^2)\lambda  z_1}{4 \pi }\right)
\end{equation}
and 
\begin{equation}
\sigma(t)^2=\sigma_0^2+\sigma_v^2 t^2
\end{equation}

The phase of the interferometer is given by a linear combination given in equation \ref{eq:Phi} of the laser phases $\bar{\phi}(z)$ at different $z$ positions.

Given the difference between the wavevectors $k_1$ and $k_2$, we make the approximation that the phase profiles at a given forward or backward position are identical for the two lasers.

Under this approximation, this rewrites as
\begin{equation}
\begin{aligned}
\Phi&=2(\phi'(A)-\phi(A))-2(\phi'(B)-\phi(B))\\
&-2(\phi'(C)-\phi(C))+2(\phi'(D)-\phi(D))
\end{aligned}
\end{equation}

Finally the averaged interferometer phase is thus given by
\begin{equation}
\bar{\Phi}=2\iint \bar{g}(x,y)\phi(x,y,0)dxdy
\label{eq:Phibar}
\end{equation}

with $G$ the Fourier transform of $g$ given by

\begin{equation}
\begin{aligned}
G(k_x,k_y)&=e^{-\frac{1}{2} k^2 \sigma_0^2} \left(\cos \left(\frac{k^2 \lambda  z_{Ai}}{4 \pi }\right)-\cos \left(\frac{k^2 \lambda  z_{Ar}}{4 \pi }\right)\right)\\
&-e^{-\frac{1}{2} k^2 \sigma_1^2} \left(\cos \left(\frac{k^2 \lambda  z_{Bi}}{4 \pi }\right)-\cos \left(\frac{k^2 \lambda  z_{Br}}{4 \pi }\right)\right)\\
&-e^{-\frac{1}{2} k^2 \sigma_1^2} \left(\cos \left(\frac{k^2 \lambda  z_{Ci}}{4 \pi }\right)-\cos \left(\frac{k^2 \lambda  z_{Cr}}{4 \pi }\right)\right)\\
&+e^{-\frac{1}{2} k^2  \sigma_2^2} \left(\cos \left(\frac{k^2 \lambda  z_{Di}}{4 \pi }\right)-\cos \left(\frac{k^2 \lambda  z_{Dr}}{4 \pi }\right)\right)
\end{aligned}
\end{equation}

with
\begin{equation}
\begin{aligned}
k^2&=k_x^2+k_y^2\\  
\sigma_1 &= \sigma(T)\\
\sigma_2 &= \sigma(2T)\\
\end{aligned}
\end{equation}

For phase fluctuations characterized by a spectrum with power spectral density (PSD) $S_\phi(k_x,k_y)$, the variance $\Sigma_{\overline{\Phi}}$ of $\overline{\Phi}$ is given by
\begin{equation}
\Sigma_{\overline{\Phi}}^2=\frac{1}{4\pi^2}\int S_\phi(k_x,k_y)\times4G(k_x,k_y)^2dk_xdk_y\\
\end{equation}

Introducing $S_\phi(k)$ the pseudo-1D PSD~\cite{Jacobs_PSD_2017} obtained after averaging over the azimuthal angle $\theta$
\begin{equation}
S_{\phi, pseudo1D}(k)=\frac{1}{2\pi}\int S_\phi(k,\theta) k d\theta
\end{equation}
we finally get
\begin{equation}
\Sigma_{\overline{\Phi}}^2=\frac{2}{\pi}\int S_{\phi, pseudo1D}(k)G(k)^2dk
\label{eq:sigmaphi}
\end{equation}

We now calculate the impact of intensity fluctuations. Even if not present in the initial plane, they develop due to the propagation of the laser beams.
Given the difference between the wavevectors $k_1$ and $k_2$, we make the approximation that the normalized intensity profiles $I_1$ and $I_2$ at a given position are identical for the two lasers.

Also we choose the intensity ratio between the two lasers so as to compensate the differential light shift, which corresponds to $\alpha+\beta=\gamma+\delta$. We denote this sum as $LS_0$. Finally, this leads to

\begin{equation*}
\delta LS = LS_0(I_1(z_{Ci}) + I_1(z_{Cr}) - I_1(z_{Bi}) - I_1(z_{Br}))
\end{equation*}

The propagation also leads to fluctuations of the amplitude $U$ of the laser field, which are related to the initial phase fluctuations, with the Fourier transform $\hat{U}$ of the field given by
\begin{equation}
\hat{U}(k_x,k_y,z)= \sin \left(\frac{k^2 \lambda  z}{4 \pi}\right)\hat{\phi}(k_x,k_y,0)
\end{equation}

Similar calculations then lead to
\begin{equation}
\Sigma_{\overline{\Phi}_{LS}}^2=\frac{2 LS_0^2 \tau_M^2}{\pi}\int S_{\phi, pseudo1D}(k)G_{LS}(k)^2dk
\label{eq:sigmaphils}
\end{equation}
with
\begin{multline}
G_{LS}(k)=e^{-\frac{1}{2} k^2 \sigma_1^2} \Biggl( \sin \left(\frac{k^2 \lambda  z_{Ci}}{ 4\pi }\right)+\sin \left(\frac{k^2 \lambda  z_{Cr}}{ 4\pi }\right)
\\ 
-\sin \left(\frac{k^2 \lambda  z_{Bi}}{4 \pi }\right)-\sin \left(\frac{k^2 \lambda  z_{Br}}{4 \pi }\right) \Biggr)
\end{multline}

\subsection{Impact of intensity fluctuations}
\label{sec:intfluct}
In the case where the profile in the initial plane features intensity fluctuations only, we get
\begin{equation}
\begin{aligned}
\hat{U}(k_x,k_y,z)&= \cos \left(\frac{k^2 \lambda  z}{4 \pi}\right)\hat{U}(k_x,k_y,0)\\
\hat{\phi}(k_x,k_y,z)&= \sin \left(\frac{k^2 \lambda  z}{4 \pi}\right)\hat{U}(k_x,k_y,0)
\end{aligned}
\end{equation}

We thus find similar expressions as \ref{eq:sigmaphi} and \ref{eq:sigmaphils}, with the cosine and sine terms accounting for the propagation of the laser fluctuations being exchanged.

\subsection{Case of the retroreflecting mirror}
\label{subsecappmirror}
Mirror surface height fluctuations lead to phase and intensity fluctuations only on the backward laser beam. The above formulas can be adapted to account for this case, by removing all terms related to the incoming beam, and by taking for $z_{Ar},z_{Cr},z_{Br},z_{Dr}$ the direct distances from the mirror to these positions, rather than the round trip distance to the mirror.


\bibliography{wfspectrum}

\end{document}